\documentclass{svjour2}                    
\smartqed  
\usepackage{graphicx}
\usepackage[latin1]{inputenc}
\usepackage{amsmath}
\usepackage{amssymb}
\newcommand{\si}[1]{\text{sgn}(}
\begin{document}

\title{A functional generalization of the field-theoretical renormalization group approach for the single-impurity Anderson model
}


\author{Hermann Freire         \and
        Eberth Corrêa 
}


\institute{Hermann Freire  \at
              Instituto de Física, Universidade Federal de Goiás, 74.001-970, Goiânia-GO, Brazil \\
              Tel.: +55 62 35211014 (Ext.250) \\
              \email{hermann@if.ufg.br}           
           \and
           Eberth Corrêa \at
              Instituto de Ciências Tecnológicas e Exatas, Universidade Federal do Triângulo Mineiro, 38.064-200, Uberaba-MG, Brazil
}

\date{Received: date / Accepted: date}

\maketitle

\begin{abstract}
We apply a functional implementation of the field-theoretical renormalization group (RG) method up to two loops to the single-impurity Anderson model. To achieve this, we follow a RG strategy similar to that proposed by Vojta \emph{et al.} [Phys. Rev. Lett. \textbf{85}, 4940 (2000)], which consists of defining a soft ultraviolet regulator in the space of Matsubara frequencies for the renormalized Green's function. Then we proceed to derive analytically and solve numerically integro-differential flow equations for the effective couplings and the quasiparticle weight of the present model, which fully treat the interplay of particle-particle and particle-hole parquet diagrams and the effect of the two-loop self-energy feedback into them. We show that our results correctly reproduce accurate numerical renormalization group data for weak to slightly moderate interactions. These results are in excellent agreement with other functional Wilsonian RG works available in the literature. Since the field-theoretical RG method turns out to be easier to implement at higher loops than the Wilsonian approach, higher-order calculations within the present approach could improve further the results for this model at stronger couplings. We argue that the present RG scheme could thus offer a possible alternative to other functional RG methods to describe electronic correlations within this model.
\keywords{Renormalization group \and Anderson impurity model \and Fermi liquid}
\PACS{71.10.Hf \and 71.10.Pm \and 71.27.+a}
\end{abstract}

\section{Introduction}

The single-impurity Anderson model (SIAM) has always been a central model in the field of strongly correlated systems \cite{Hewson}. It was originally proposed to describe microscopically the effect of a localized impurity embedded in a metallic host on its conduction electrons. Recently, the importance of this model has increased even further. Thanks to ground-breaking advances in nanotechnology, it is now possible to simulate experimentally this model in nanostructure devices such as quantum dots coupled to metallic leads \cite{Goldhaber}. This has opened a new arena for exploration of correlation effects in many-body systems. On the theoretical side, this renewed interest in the model is also related to the discovery that the paradigmatic Hubbard model in the appropriate limit of infinite spatial dimensions\cite{Metzner} (or infinite coordination number)  can in fact be mapped onto an effective SIAM with an additional self-consistency condition \cite{Kotliar}. This discovery led to the development of the Dynamical Mean-Field Theory (DMFT) which, combined with first-principle methods such as the Density Functional Theory \cite{Kohn} to determine the electronic structure, is currently being applied to realistic strongly correlated materials with good success.

However, since it is difficult to solve the SIAM in its most general form (except for some special conditions where it can be solved with the help of the Bethe ansatz \cite{Tsvelick} or the Numerical Renormalization Group (NRG) \cite{Wilson}), this represents nowadays an obstacle to some extent for the systematic application of DMFT to many strongly-correlated materials. There is a lack of reliable methods to study some of the complicated generalizations of the SIAM generated by DMFT calculations, especially in the strong-coupling regime. To tackle this problem, it is important to set up new theoretical schemes which offer the advantage of being flexible in the implementation or generalization and which require preferably low computational effort. In this respect, we mention that a major step forward has been recently achieved via diagrammatic quantum Monte Carlo which turns out to be free from the fermionic sign problem \cite{Werner}.

Another promising semi-analytical method developed in recent years is based on the functional generalization of the renormalization group (RG) approach. Several different implementations of this technique are already available in the literature, both in the Wilsonian version \cite{Wetterich,Morris,Shankar,Salmhofer,Metzner2,Zanchi,Honerkamp,Kopietz,Tsai} and the field-theoretical version \cite{Yakovenko,Alvarez,Doucot,Freire,Edwards}. They have been applied to many electronic models ranging from the two-dimensional (2D) Hubbard model to quantum impurity models such as the SIAM or the closely related Kondo model. Concerning the 2D Hubbard model, for instance, some of these functional RG works successfully reproduced within a static approximation both at one-loop \cite{Metzner2,Honerkamp,Zanchi,Alvarez} and two-loop \cite{Freire2,Katanin} levels an antiferromagnetic phase in the model near half-filling and the onset of a $d_{x^{2}-y^{2}}$-wave singlet superconducting phase away from half-filling for weak-to-moderate couplings. This agrees qualitatively with the physics displayed by the cuprate superconductors and gives further support to the point of view that this 2D model might indeed capture some important features of these strongly-correlated materials.

Given these encouraging results, several functional RG studies have also been initiated nowadays to discuss quantum impurity problems \cite{Meden,Bartosch} such as the SIAM with the important goal of reproducing some key aspects of this model, most notably, its dynamical properties for all interaction strengths and also the correct emergence of the exponential Kondo scale at strong coupling. In addition to providing useful benchmarks to the functional RG scheme -- more specifically, to the most common approximations used in these approaches -- these studies could also provide alternative and computationally amenable semi-analytical quantum impurity solvers which, as we have seen before, are a central ingredient in the context of DMFT calculations.

In this respect, Karrasch \emph{et al.} implemented a Wilsonian functional RG study at one-loop level \cite{Meden} of several important quantities of the model such as the effective mass, the static spin susceptibility and the spectral function. They concluded that while their approach yielded very good results for weak couplings, they still could not reproduce the exponentially small Kondo asymptotics at higher couplings. Besides that, due to difficulties (ill-posed problem) with the stability of the analytic continuation of their functional RG data for the the
single-particle Green's function, they were not able to establish conclusively the formation of the so-called Hubbard satellite bands which emerge at high energies in the impurity spectral function for strong coupling. Recently, two different works reported some progress on these problems: Using a Hubbard-Stratonovich decoupling of the exact Fermi-Bose RG equations, Isidori \emph{et al.} managed to obtain a three-peak structure in the spectral function of the model at finite temperatures \cite{Isidori} in qualitative agreement with precise NRG calculations for intermediate couplings. Also, Jakobs \emph{et al.} applying a functional RG scheme within the Keldysh formalism \cite{Schoeller} obtained qualitative similar results for the SIAM in the equilibrium situation for weak-to-moderate interaction. Despite that, both approaches could not obtain an exponential narrowing of the quasiparticle peak at strong coupling, which suggests that the Kondo scale emergence is still not completely included at the level of approximation used in those works. In this sense, a next logical step would be to check if adding higher-order terms would improve further these results for the model.
It is interesting to mention that higher-order terms are also recognized as an important contribution in order to compute analytic results for transport quantities and dynamical correlation functions of the Kondo model out
of equilibrium within the so-called real-time renormalization group method in frequency space \cite{Schoeller2,Schuricht,Andergassen}.

In the present work, we implement a two-loop calculation of the the self-energy together with a calculation of the the renormalized coupling functions for the SIAM at zero temperature using a functional generalization of the standard field-theoretical RG approach. This analysis is important in view of the fact that it includes self-energy effects in all coupled integro-differential RG flow equations. In addition to this, it also includes, by construction, the dynamics of all frequency-dependent quantities in the model. Since the functional field-theoretical RG is not widely
known among condensed matter physicists we present the method at length and in full detail. As an initial test for this RG scheme, we present both analytical and numerical results for the renormalization of the effective coupling functions and the quasiparticle weight in the low-energy limit for all interaction strengths. Afterwards, we proceed to benchmark our results against exact (or highly accurate) data for this model such as the Bethe ansatz and NRG.

This paper is structured as follows. In Sec. II, the single impurity Anderson model that we want to investigate is introduced. Next (in Sec. III), we explain the functional field-theoretical RG methodology used to study the model. The two-loop RG flow equations are derived explicitly in Sec. IV. We then proceed to solve numerically the coupled integro-differential RG equations and discuss the main results in Sec. V. Finally, we present our conclusions from this study.

\section{Model}

The Hamiltonian of the SIAM is given by

\vspace{-0.2cm}

\begin{eqnarray}
 \hat{H} & = & \sum_{ \mathbf{k} \sigma} \epsilon_{ \mathbf{k} }
 \hat{c}^{\dagger}_{ \mathbf{k} \sigma } \hat{c}_{ \mathbf{k} \sigma }+ \sum_{\sigma} E_d \hat{d}^{\dagger}_{\sigma} \hat{d}_{\sigma}
 \nonumber
\\
 & + &
 \sum_{\mathbf{k} \sigma} ( V_{\mathbf{k}}^{\ast}
 \hat{d}^{\dagger}_{ \sigma} \hat{c}_{\mathbf{k} \sigma } + V_{\mathbf{k}}
 \hat{c}^{\dagger}_{\mathbf{k} \sigma } \hat{d}_{\sigma} )+U \hat{d}^{\dagger}_{\uparrow} \hat{d}_{\uparrow}
 \hat{d}^{\dagger}_{\downarrow} \hat{d}_{\downarrow},
 \label{eq:Hdef}
\end{eqnarray}

\noindent where $\hat{c}_{\mathbf{k} \sigma}$ annihilates a noninteracting conduction electron with
momentum ${\mathbf{k}}$, energy dispersion $\epsilon_{ \mathbf{k} }$, and spin projection $\sigma$,
whereas $\hat{d}_{\sigma}$ annihilates
a localized electron with
energy $E_d$ and on-site repulsion $U$.
The hybridization between
the localized electrons and the conduction electrons is characterized by the
amplitude $V_{\mathbf{k}}$.

If we use a coherent-state functional integral representation of Eq. (\ref{eq:Hdef}) at constant chemical potential $\mu$ and integrate out the conduction electrons, the corresponding model at $T=0$ becomes readily described by the action

\vspace{-0.2cm}

\begin{eqnarray}\label{action}
&&S[d,\bar{d}]=-\sum_{\sigma}\int_{\omega}[i\omega-\xi_{0}-\Delta(i\omega)]\bar{d}_{\sigma}(\omega)d_{\sigma}(\omega)\nonumber\\
&&+U\int_{\omega_1,\omega_2,\omega_3}\bar{d}_{\uparrow}(\omega_{1}+\omega_{2}-\omega_{3})\bar{d}_{\downarrow}(\omega_3)d_{\downarrow}(\omega_2)
d_{\uparrow}(\omega_1),\nonumber\\
\end{eqnarray}

\noindent where $\xi_{0}=E_d -\mu$ and $\int_\omega= \int\frac{d\omega}{2\pi}$. In addition, $\bar{d}_{\sigma}$ and $d_{\sigma}$ are, respectively, the creation and annihilation Grassmann fields for localized electrons with spin projection $\sigma$ and with the hybridization function given by $\Delta(i\omega) =\sum_{ \mathbf{k}}  \frac{ | V_{\mathbf{k}} |^2}{ i \omega - \epsilon_{\mathbf{k}}}$. The above action then defines our bare quantum field theory which must be appropriately regularized in the UV regime (more details on the regularization procedure in the next section). For simplicity, we shall consider here the SIAM in the so-called wide-band limit such that $\Delta(i\omega)=-i\Delta\text{sgn}(\omega)$, where $\Delta$ is a constant. It is also convenient to make the following choice of variables, i.e. $\xi_0=-U/2$. When this is done, the model becomes completely particle-hole symmetric.

If we treat the interaction of Eq. (\ref{action}) within the self-consistent Hartree-Fock approximation, we obtain the
renormalized excitation energy $\xi(\omega)=\xi_{0}+\Sigma_{HF}(\omega)$, where

\vspace{-0.2cm}

\begin{equation}\label{HF}
\Sigma_{HF}(\omega)=U\int_{\omega'}G_{HF}(i\omega-i\omega'),
\end{equation}

\noindent with $G_{HF}(i\omega)=1/[i\omega-\xi(\omega)+i\Delta\text{sgn}(\omega)]$ being the self-consistent Hartree-Fock impurity Green's function of the model. Since, for frequency-independent interaction, one obtains that $\Sigma_{HF}(\omega)=U/2$, this latter contribution precisely cancels $\xi_0$ in the particle-hole symmetric case. However, as will become clear soon, despite the fact that the initial interaction of the SIAM is indeed a constant, it will acquire a frequency-dependent structure under the RG transformation. For this reason, in order to preserve the particle-hole symmetry of the model during the whole RG flow, we must go back to Eq. (\ref{action}) and define a frequency-dependent impurity energy $\xi_0=\xi_{0}(\omega)$, such that the condition $\xi_0(\omega)+\int_{\omega'}U(\omega,\omega',\omega')G_{HF}(i\omega-i\omega')=0$ is always enforced for this case. As a consequence of this, the corresponding Hartree-Fock impurity propagator will naturally take the standard result

\vspace{-0.2cm}

\begin{equation}\label{HFpropagator}
G_{HF}(\omega)=\frac{1}{i\omega+i\Delta\text{sgn}(\omega)}.
\end{equation}

\section{RG Methodology}

The outline of the field-theoretical RG method can be generally summarized in the following way (for more details, see, e.g., Ref. \cite{Peskin}). Typically, in correlated systems if one applies a naive perturbation theory for a given electronic model, divergences or non-analyticities often appear in the low-energy limit at the calculation of several physical quantities of the model such as, e.g., irreducible two-particle vertices, the self-energy and also several order-parameter susceptibilities. This result usually implies that the perturbation theory setup is not appropriately formulated. The field-theoretical RG approach circumvents this problem by means of a reorganization of the terms in the conventional perturbative approach. More precisely, by rewriting the experimentally unobserved bare quantities of the microscopic field theory model (such as, e.g., the fermionic fields and the coupling parameters) in terms of the corresponding  renormalized parameters, it is possible to construct a new renormalized perturbation theory which is, by construction, convergent in the low-energy limit. The bare quantities are always defined at the ultraviolet (UV) scale of the model, whereas the renormalized parameters are conveniently defined at a physically measurable low-energy scale. Since the renormalized parameters are the ones that are physically probed from the experimental point of view, they must come out necessarily UV cutoff-independent. This regularization procedure has to be implemented order by order in perturbation theory. If this program is successfully accomplished, then the field theory model is said to be properly renormalized.

It is interesting to note however that, unlike the situation for many lattice models such as, e.g, the 2D Hubbard model, where divergences and non-analyticities in Feynman diagrams often abound, perturbation theory is regular for the SIAM. Therefore, one might naively think, at first sight, that there should be no intrinsic need in this model for a RG-based theoretical attack. This view is overly restrictive though, especially due to the fact that the emergence of an exponentially dependent (in $U$) Kondo scale at the model indeed motivates many-body resummations of certain classes of diagrams. The RG framework is an ideal tool for achieving this goal in an unbiased manner.

In this work, we implement a functional field-theoretical RG approach up to two-loops for the SIAM in order to calculate the dynamics of all frequency-dependent renormalized quantities of the model. This analysis is important for the following reasons: Firstly, despite the fact that the frequency dependencies of the renormalized parameters could be classified as irrelevant in a power counting sense, they do have a crucial effect on the low-energy behavior of the model especially at stronger couplings. Secondly, in order to evaluate experimentally relevant dynamical quantities such as the spectral density function and the dynamical susceptibilities of the model, it is essential to include such a generalization from the outset in all physical parameters of the RG approach. Lastly, we briefly mention that finite-energy properties of the model are also important if one is willing to extend the functional field-theoretical RG approach to the non-equilibrium situation, e.g., to calculate transport properties in quantum dots (see, for instance, Ref. \cite{Schoeller} for a Wilsonian RG approach to this problem).

\begin{figure}[t]
  \includegraphics[width=1.9in]{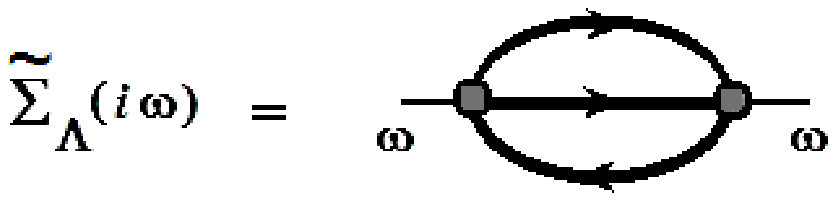}\\
  \caption{The self-energy diagram of the SIAM at two-loop order. The thick lines represent the interacting Green's function of the model.}\label{Selfenergy}
\end{figure}

To begin our analysis, we describe the underlying strategy of our RG scheme. We shall follow a RG method that has some similarities with the approach proposed by Vojta \emph{et al.} in a different context \cite{Vojta}, namely to discuss possible critical points in high-T$_{c}$ superconductors. In our case, in order to implement a functional generalization of the field-theoretical RG method and also to avoid possible artificial non-analyticities generated by the regularization procedure in the SIAM, it is better not to impose a sharp UV cutoff in the field-theory model defined by Eq. (\ref{action}). For this reason, we define a soft regulator $K({|\omega|}/{\Lambda})$ in the space of Matsubara frequencies for the fully interacting Green's function (where $\Lambda$ stands for the UV cutoff of the model). The exact impurity propagator should then be given by

\vspace{-0.2cm}

\begin{equation}\label{1}
G_{\Lambda}(i\omega)=\frac{1}{i\omega+i\Delta\text{sgn}(\omega)-\widetilde{\Sigma}_{\Lambda}(\omega)}K\left({|\omega|}/{\Lambda}\right),
\end{equation}

\noindent where $\widetilde{\Sigma}_{\Lambda}(\omega)=\Sigma_{\Lambda}(\omega)-\Sigma_{\Lambda}^{HF}(\omega)$ is the self-energy of the SIAM without the Hartree-Fock term, since this latter contribution is already included in the ``noninteracting'' propagator of the model as we have shown previously. In addition to this, $K(y)$ must be some smooth decaying function which must satisfy $K(0)=1$. Following Ref. \cite{Vojta}, we conveniently choose $K(y)=e^{-y}$. This choice of the UV regulator turns out also to be more amenable from both analytical and numerical point of view. (We parenthetically note here that a sharp cutoff would correspond to a Heaviside step function $K(y)=\theta(y-1)$.) As we will discuss shortly, to derive the RG flow equations, in which degrees of freedom with Matsubara frequencies lying between $\Lambda$ and $\Lambda-d\Lambda$ are successively integrated out, we must take a $\Lambda\,(d/d\Lambda)$ derivative of all irreducible two-particle vertices and the self-energy of the model.

Since we will be interested in the low-energy behavior of the SIAM -- particularly in the possible emergence of the dynamically generated Kondo scale -- we shall focus here on the low-frequency range $|\omega|\lesssim \Delta$ of the model. In this regime, it is reasonable to assume that the renormalized Green's function is given by the low-frequency Fermi liquid form

\vspace{-0.4cm}

\begin{equation}\label{1a}
G_{\Lambda}(i\omega)=\frac{Z_{\Lambda}}{i\omega+iZ_{\Lambda}\Delta\text{sgn}(\omega)}K\left({|\omega|}/{\Lambda}\right),
\end{equation}

\noindent i.e. the self-energy of the model is given by $\widetilde{\Sigma}_{\Lambda}(i\omega)=i\omega(1-1/Z_{\Lambda})$, where $Z_{\Lambda}=(1-\partial\widetilde{\Sigma}_{\Lambda}(i\omega)/\partial(i\omega)|_{\omega=0})^{-1}$ is the quasiparticle weight. At weak-coupling, we know from second-order perturbation theory \cite{Yamada} that this quantity should be given by

\vspace{-0.4cm}

\begin{equation}\label{PT}
Z_{\Lambda\rightarrow 0}= 1+ \left(3-\frac{\pi^2}{4}\right)(U/\pi\Delta)^2 + ...,
\end{equation}

\noindent whereas in the strong-coupling limit $U\rightarrow \infty$, from Bethe ansatz results \cite{Tsvelick}, the quasiparticle weight is given asymptotically by

\vspace{-0.4cm}

\begin{equation}\label{1b}
Z_{\Lambda\rightarrow 0}\sim \sqrt{\frac{8\,U}{\pi^2\Delta}}\exp[-\pi\,U/8\Delta].
\end{equation}

In what follows, we calculate explicitly the RG flow equations up to two-loop order for the SIAM. In order to do this, we choose a parametrization of the interaction parameter of the model in a way which manifestly incorporates the energy and the spin conservation

\vspace{-0.4cm}

\begin{eqnarray}\label{2}
&&U_{\Lambda}(\omega_1 \sigma_1,\omega_2 \sigma_2,\omega_3 \sigma_3,\omega_4 \sigma_4)=\delta(\omega_1+\omega_2-\omega_3-\omega_4)
\bigg\{U_{\uparrow}^{\Lambda}(\omega_1,\omega_2,\omega_3)[\delta_{\sigma_1\uparrow}\delta_{\sigma_2\uparrow}\delta_{\sigma_3\uparrow}
\delta_{\sigma_4\uparrow}+\delta_{\sigma_1\downarrow}\delta_{\sigma_2\downarrow}\delta_{\sigma_3\downarrow}
\delta_{\sigma_4\downarrow}]\nonumber\\
&&+U_{\uparrow\downarrow}^{\Lambda}(\omega_1,\omega_2,\omega_3)[\delta_{\sigma_1\uparrow}\delta_{\sigma_2\downarrow}\delta_{\sigma_3\uparrow}
\delta_{\sigma_4\downarrow}+\delta_{\sigma_1\downarrow}\delta_{\sigma_2\uparrow}\delta_{\sigma_3\downarrow}
\delta_{\sigma_4\uparrow}-\delta_{\sigma_1\uparrow}\delta_{\sigma_2\downarrow}\delta_{\sigma_3\downarrow}
\delta_{\sigma_4\uparrow}-\delta_{\sigma_1\downarrow}\delta_{\sigma_2\uparrow}\delta_{\sigma_3\uparrow}
\delta_{\sigma_4\downarrow}]\bigg\}.
\end{eqnarray}

The RG methodology we apply now is standard and follows closely the field-theoretical method (see, e.g., Refs. \cite{Peskin,Vojta}). We begin by first calculating the RG flow equation for the self-energy of the model at two-loop order as depicted in Fig. 1. As a result, we obtain

\vspace{-0.2cm}

\begin{eqnarray}\label{3}
\partial_\Lambda\widetilde{\Sigma}_\Lambda (i\omega)&=&\frac{i Z_{\Lambda}^3}{4\pi^2\Lambda^2}\int_{-\infty}^{+\infty}d\omega'\int_{-\infty}^{+\infty}d\omega''
\Big[U^{\Lambda}_{\uparrow}(\omega,\omega+\omega''-\omega',\omega'')
U^{\Lambda}_{\uparrow}(\omega',\omega'',\omega+\omega''-\omega')\nonumber\\
&+&2U^{\Lambda}_{\uparrow\downarrow}(\omega,\omega+\omega''-\omega',\omega'')
U^{\Lambda}_{\uparrow\downarrow}(\omega',\omega'',\omega+\omega''-\omega')
-U^{\Lambda}_{\uparrow}(\omega,\omega'',\omega')
U^{\Lambda}_{\uparrow}(\omega+\omega''-\omega',\omega',\omega)\nonumber\\
&-&2U^{\Lambda}_{\uparrow\downarrow}(\omega+\omega''-\omega',\omega',\omega)
U^{\Lambda}_{\uparrow\downarrow}(\omega,\omega'',\omega')\Big]
K\left({|\omega'|}/{\Lambda}\right)K\left({|\omega''|}/{\Lambda}\right)K\left({|\omega+\omega''-\omega'|}/{\Lambda}\right)\nonumber\\
&\times &\frac{(|\omega'|+|\omega''|+|\omega+\omega''-\omega'|)}{[\omega'+Z_{\Lambda}\Delta\text{sgn}(\omega')][\omega''+Z_{\Lambda}\Delta\text{sgn}(\omega'')]
[\omega+\omega''-\omega'+Z_{\Lambda}\Delta\text{sgn}(\omega+\omega''-\omega')]}\,.\\ \nonumber
\end{eqnarray}

\noindent An important point we wish to emphasize here is that it is essential to use the interacting Green's function (i.e., which includes self-energy feedback) in the RG perturbative scheme instead of the Hartree-Fock Green's function in order to remove the unphysical Stoner instability in the model (for a discussion on this issue, see, e.g., Ref. \cite{Bartosch}). If we now use the many-body definition displayed above for the quasiparticle weight factor $Z_{\Lambda}$, we can straightforwardly derive the RG flow equation
at two-loop order for this quantity as follows

\vspace{-0.2cm}

\begin{equation}\label{4}
\Lambda\frac{d\,\ln Z_{\Lambda}}{d\Lambda}=\eta,
\end{equation}

\noindent where $\eta=\Lambda Z_{\Lambda}\frac{\partial}{\partial(i\omega)}\big[\partial_{\Lambda} \widetilde{\Sigma}_{\Lambda}(i\omega)\big]_{\omega=0}$. In this way, $\eta$ is given by

\begin{eqnarray}\label{5}
\eta&=&\frac{Z_{\Lambda}^4}{4\pi^2\Lambda}\int_{-\infty}^{\infty} d\omega'\int_{-\infty}^{\infty} d\omega''
\Big\{\big[U_{\uparrow}^{\Lambda}(\omega,\omega+\omega''-\omega',\omega'')
U_{\uparrow}^{\Lambda}(\omega',\omega'',\omega+\omega''-\omega')\nonumber\\
&+&2U_{\uparrow\downarrow}^{\Lambda}(\omega,\omega+\omega''-\omega',\omega'')
U_{\uparrow\downarrow}^{\Lambda}(\omega',\omega'',\omega+\omega''-\omega')
-U_{\uparrow}^{\Lambda}(\omega,\omega'',\omega')
U_{\uparrow}^{\Lambda}(\omega+\omega''-\omega',\omega',\omega)\nonumber\\
&-&2U_{\uparrow\downarrow}^{\Lambda}(\omega+\omega''-\omega',\omega',\omega)
U_{\uparrow\downarrow}^{\Lambda}(\omega,\omega'',\omega')
\big]\mathcal{F}_{\Lambda}(\omega,\omega',\omega'')\Big\}_{\omega=0},
\end{eqnarray}

\noindent where

\vspace{-0.1cm}

\begin{eqnarray}\label{6}
\mathcal{F}_{\Lambda}(\omega,\omega',\omega'')&=&\frac{K(|\omega'|/\Lambda)K(|\omega''|/\Lambda)K(|\omega+\omega''-\omega'|/\Lambda)}
{(\omega'+Z_{\Lambda}\Delta\,\text{sgn}(\omega'))(\omega''+Z_{\Lambda}\Delta\,\text{sgn}(\omega''))(\omega+\omega''-\omega'+Z_{\Lambda}\Delta\,\text{sgn}(\omega+\omega''-\omega'))}\nonumber\\
&\times&\Big[\frac{d}{d\omega}|\omega+\omega''-\omega'|-\frac{(|\omega'|+|\omega''|+|\omega+\omega''-\omega'|)}{\Lambda}
\frac{d}{d\omega}|\omega+\omega''-\omega'|\nonumber\\
&-&\frac{(|\omega'|+|\omega''|+|\omega+\omega''-\omega'|)(1+Z_{\Lambda}\Delta\frac{d}{d\omega}\,\text{sgn}(\omega+\omega''-\omega'))}
{(\omega+\omega''-\omega'+Z_{\Lambda}\Delta\,\text{sgn}(\omega+\omega''-\omega'))}\Big].\\ \nonumber
\end{eqnarray}

\begin{figure}[t]
  \includegraphics[width=3.2in]{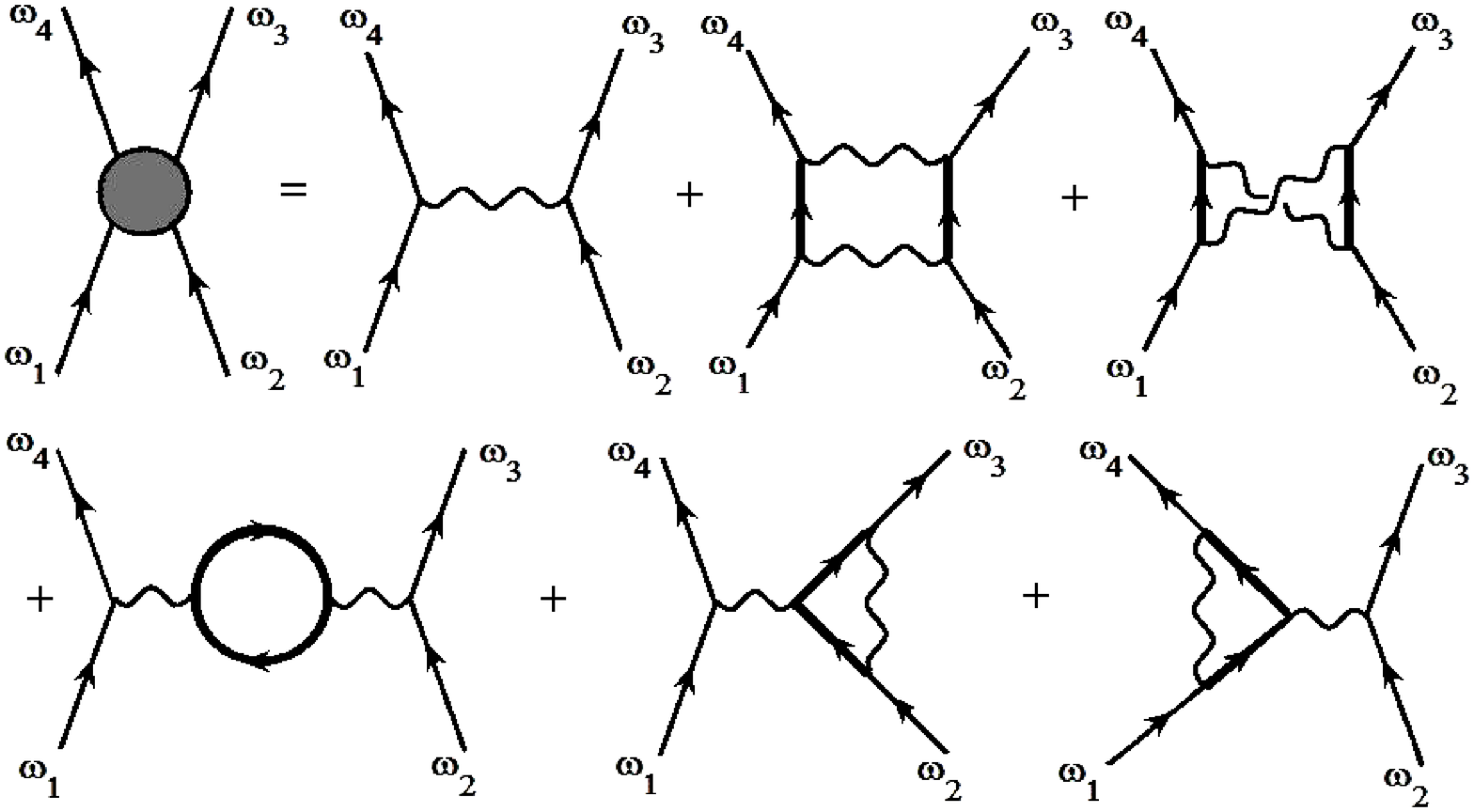}\\
  \caption{The parquet diagrams for the renormalized irreducible two-particle vertices of the SIAM  with two-loop self-energy feedback including the competition of both particle-particle and particle-hole contributions. The thick lines represent the interacting Green's function of the model.}\label{parquet}
\end{figure}

Next, we proceed to calculate the RG flow equations for the coupling functions of the SIAM. Since all the renormalized parameters are the physically measurable quantities, they do not depend on the UV cutoff scale $\Lambda$. Therefore, we must set $\Lambda\, d\Gamma_{R,i}(\omega_{1},\omega_{2},\omega_{3})/d\Lambda=0$ for $i=\,\uparrow\text{and}\uparrow\downarrow$, where $\Gamma_{R,i}(\omega_{1},\omega_{2},\omega_{3})$ denote the renormalized irreducible two-particle vertices \cite{Shankar}. The corresponding Feynman diagrams for the coupling functions with two-loop self-energy feedback are schematically shown in Fig. 2. As a result, we obtain the following coupled integro-differential RG equations which fully describe the interplay of both particle-particle and particle-hole vertex corrections in the model (the so-called parquet diagrams), i.e.

\begin{equation}\label{7}
\Lambda\frac{dU^{\Lambda}_{\uparrow}(\omega_{1},\omega_{2},\omega_{3})}{d\Lambda}=
A^{\Lambda}_{\uparrow}(\omega_{1},\omega_{2},\omega_{3})
+B^{\Lambda}_{\uparrow}(\omega_{1},\omega_{2},\omega_{3})+C^{\Lambda}_{\uparrow}(\omega_{1},\omega_{2},\omega_{3}),
\end{equation}

\noindent and

\begin{equation}\label{8}
\Lambda\frac{dU^{\Lambda}_{\uparrow\downarrow}(\omega_{1},\omega_{2},\omega_{3})}{d\Lambda}=
A^{\Lambda}_{\uparrow\downarrow}(\omega_{1},\omega_{2},\omega_{3})+B^{\Lambda}_{\uparrow\downarrow}(\omega_{1},\omega_{2},\omega_{3})
+C^{\Lambda}_{\uparrow\downarrow}(\omega_{1},\omega_{2},\omega_{3}),
\end{equation}

\vspace{+0.6cm}

\noindent where $\omega_4=\omega_1+\omega_2-\omega_3$ and the functions $A^{\Lambda}_{\uparrow}$, $B^{\Lambda}_{\uparrow}$, $C^{\Lambda}_{\uparrow}$ $A^{\Lambda}_{\uparrow\downarrow}$, $B^{\Lambda}_{\uparrow\downarrow}$, and $C^{\Lambda}_{\uparrow\downarrow}$ are given explicitly in Appendix A. The Eqs. (\ref{4}), (\ref{7}), and (\ref{8}) must be solved simultaneously in order to analyze the low-energy behavior of this model. Since it is impossible to obtain an analytical solution for these equations in this case, we must then resort to numerical methods. This will be done in the next section.

\section{Numerical Results}

We now devote this section to the numerical analysis of the integro-differential RG equations up to two-loop order derived above. Following previous works \cite{Meden,Isidori}, since we are mostly interested in the low-energy physics of the SIAM in order to reproduce the Kondo scale we discretize the Matsubara frequencies by introducing the geometric mesh

\vspace{-0.2cm}

\begin{equation}\label{mesh}
\omega_n=\pm\omega_0\frac{a^n -1}{a-1} \hspace{1cm} n=1...N,
\end{equation}

\noindent where the free adjustable parameters $\omega_0>0$ and $a>1$ determine the spacing of the discretized frequencies and $N$ is the total number of frequencies to be used. For this discretization procedure, the low-frequency regime will be better resolved than the corresponding high-frequency sector, which is justified since the Kondo peak emerges at $\omega=0$ in the model. A typical choice of these parameters is, e.g.,

\vspace{-0.2cm}

\begin{equation}
\omega_0=10^{-5}\Delta, \hspace{0.2cm} a=3.5\nonumber, \hspace{0.2cm}\Lambda_0=10\Delta,\hspace{0.2cm}N=30\hspace{0.1cm}\text{points},
\end{equation}

\noindent which allows one to solve the flow equations at two-loop order with relatively low computational effort. Despite this moderate choice of number of points in the discretization procedure, we
point out that our results show good convergence properties
in the low-energy limit. This choice implies that we have to keep track of the RG flow of thousands of couplings altogether. Since the coupling functions in the RG equations need to be evaluated for arbitrary arguments -- which in general do not coincide with a point of the mesh -- an interpolation procedure has to be implemented. In this work, we shall use a linear interpolation. As a result of this, all the integrals in the flow equations will be solved by using standard Gaussian quadrature routines. Afterwards, the resulting set of coupled differential RG equations are then solved by means of high-precision fourth-order Runge-Kutta method.

\begin{figure}[t]
  \includegraphics[width=2.3in]{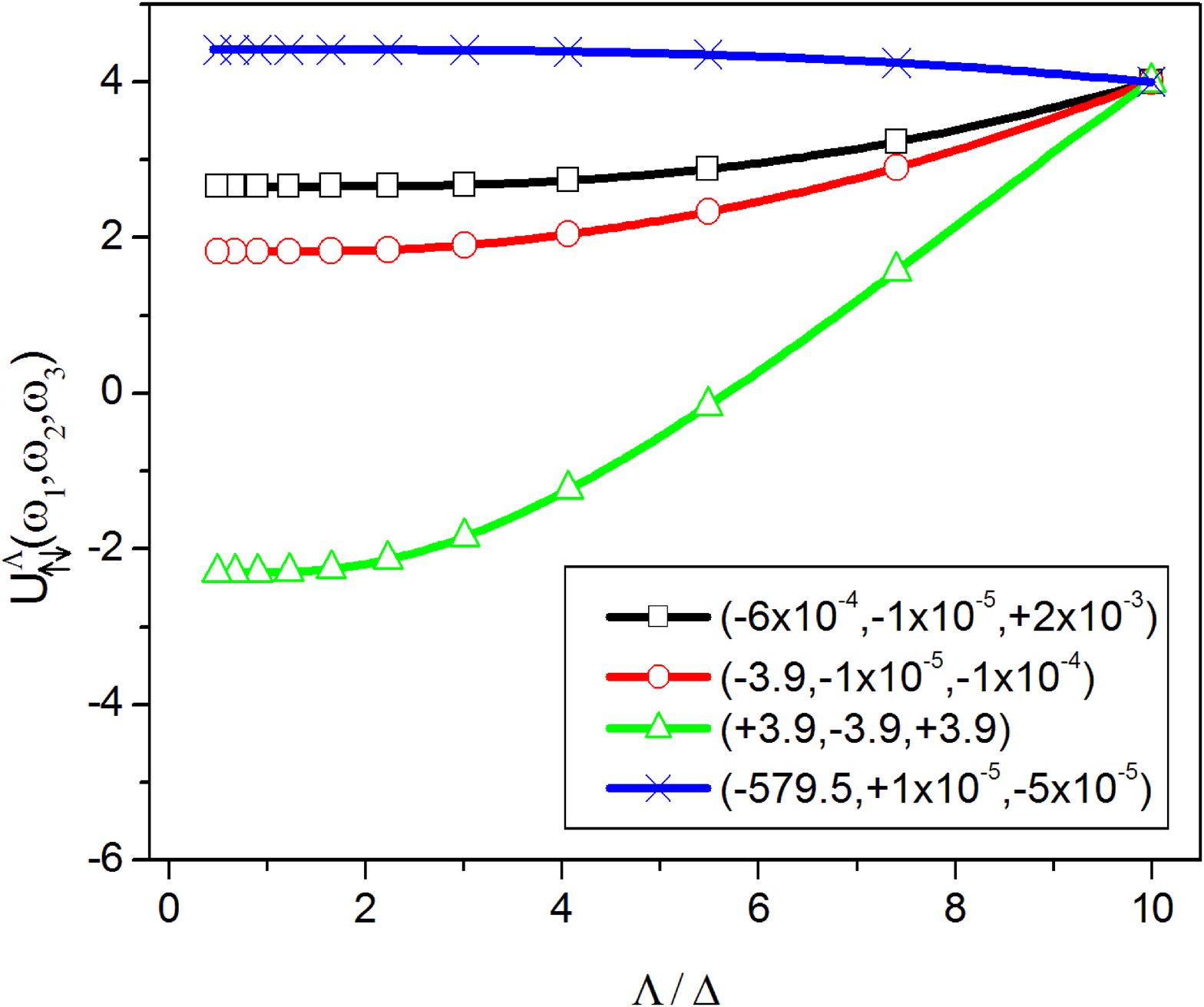}\includegraphics[width=2.3in]{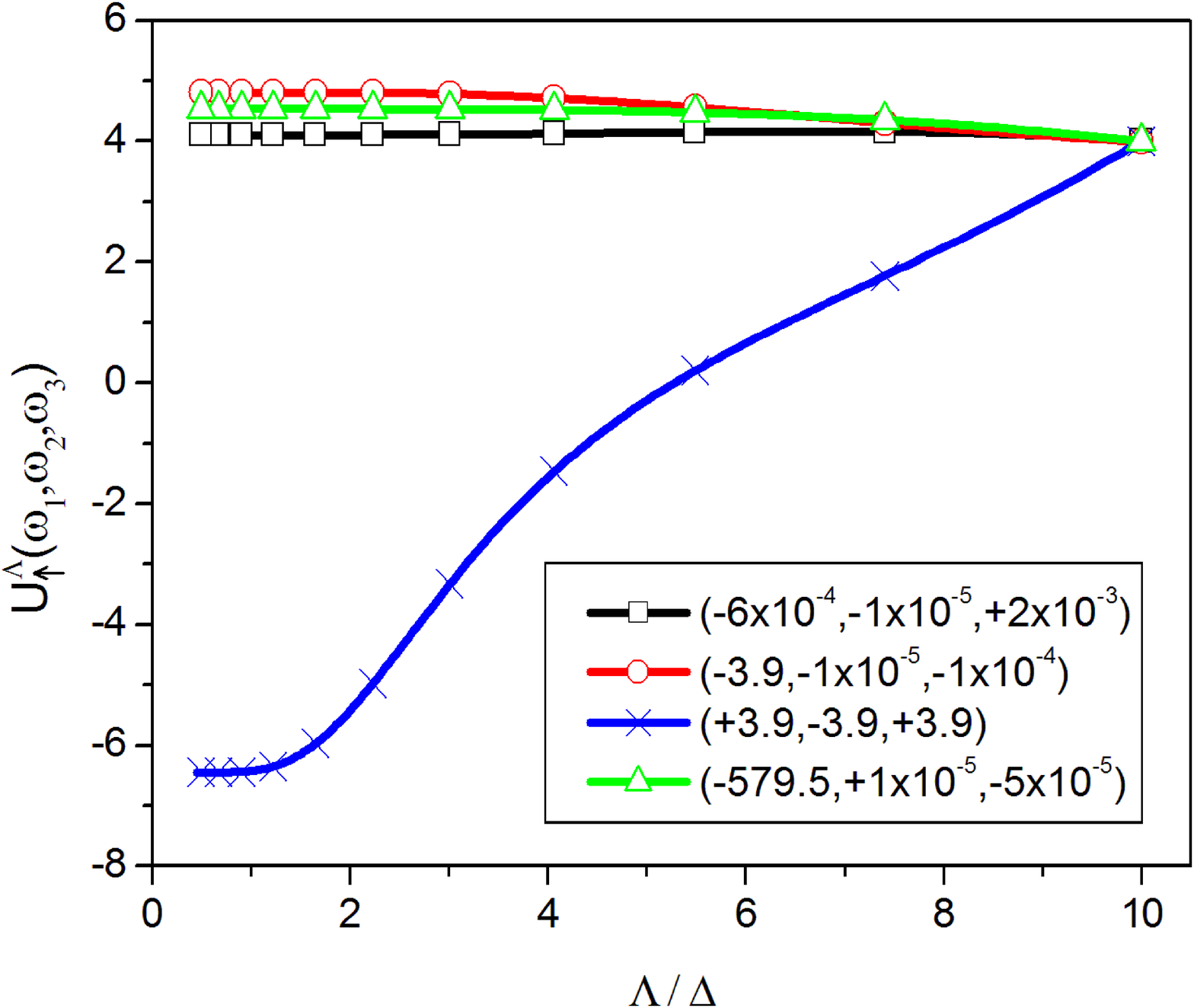}\\
  \caption{(Color online) RG flow with two-loop self-energy corrections for some choices of external frequencies of the couplings $U^{\Lambda}_{\uparrow\downarrow}(\omega_{1},\omega_{2},\omega_{3})$ and $U^{\Lambda}_{\uparrow}(\omega_{1},\omega_{2},\omega_{3})$ for the SIAM (in units of $\Delta)$ as a function of the ratio $\Lambda/\Delta$. The bare coupling at the initial cutoff $(\Lambda_0/\Delta)=10$ was chosen to be $(U/\Delta)=4$.}\label{flow1}
\end{figure}

\begin{figure}[b]
  \includegraphics[width=2.8in]{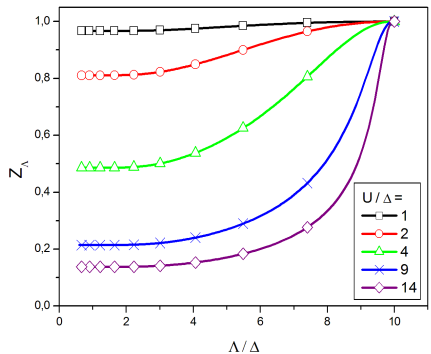}\\
  \caption{(Color online) RG flow of the quasiparticle weight $Z_{\Lambda}$ as a function of $\Lambda/\Delta$ for several choices of initial interaction strength $U$ within the functional field-theoretical renormalization group up to two-loop order.}\label{flow2}
\end{figure}

First, we focus our attention on the numerical solution of Eqs. (\ref{7}) and (\ref{8}) for the renormalized couplings as a
function of $\Lambda$. The choice of the initial conditions at $\Lambda = \Lambda_0$ in the RG equations are given by $U_{\uparrow}=U_{\uparrow\downarrow}=U$. As an illustrative example, we choose the initial interaction strength $(U/\Delta) = 4$. Our results for this choice of interaction are displayed in Fig. 3. In this plot, even though the couplings are initially constant, we demonstrate that they acquire as expected a
distinct frequency structure as we integrate out degrees of freedom towards the low-energy limit (i.e. $\Lambda\rightarrow 0$). Technically speaking, this happens because of the interplay of parquet particle-particle and particle-hole classes of diagrams in the vertex corrections which are included to infinite order in our field-theoretical RG scheme. In other words, despite the irrelevance in a power counting sense of the frequency dependence of the coupling functions for the SIAM, they do play an important role in the corresponding low-energy dynamics for this case.

\begin{figure}[t]
  \includegraphics[width=2.8in]{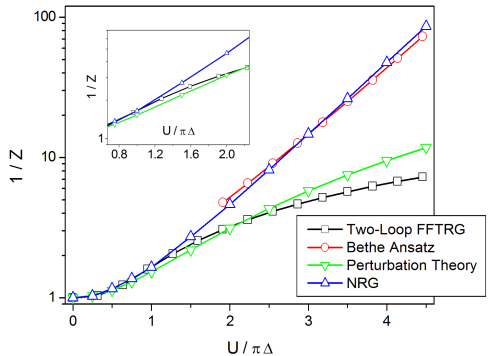}\\
  \caption{(Color online) Inverse quasiparticle weight $Z_{\Lambda}^{-1}$ in the limit $\Lambda\rightarrow 0$ as a function of $U/\pi\Delta$ within the functional field-theoretical renormalization group (FFTRG) up to two-loop order. For comparison, we have displayed highly accurate NRG data obtained in Ref.\cite{Isidori} and second-order perturbation theory result from Eq. (\ref{PT}). In addition, we also plot the Bethe ansatz result of Eq. (\ref{1b}) for $(U/\pi\Delta)\gtrsim 2$, even though this latter expression is only valid in the limit $U\rightarrow \infty$.}\label{plot}
\end{figure}

Next we present the results of the RG flow for the quasiparticle weight $Z_{\Lambda}$ of the model as a function of the ratio $\Lambda/\Delta$. Since at the UV cutoff scale ($\Lambda=\Lambda_0$) there should be no quantum fluctuations included in the RG scheme, the quasiparticle weight must be initially equal to unity (i.e. it must assume its noninteracting value). Naturally, the final value of $Z_{\Lambda}$ in the low-energy limit will depend crucially upon the bare interaction strength $U$. Our results are displayed in Fig. 4 for some choices of initial interaction. In this figure, we observe clearly a strong suppression of $Z_{\Lambda\rightarrow 0}$ with increasing $U$. Despite that, the quasiparticle weight always approach a finite value in the low-energy limit. In other words, there is no sign of instability or artificial spin symmetry breaking during the RG flow within our scheme. This is consistent with the fact that the SIAM conforms to a Fermi liquid description for all couplings at low-energies. Our RG approach therefore captures successfully this important aspect of the problem.

In order to verify if the Kondo scale of the model is reproduced by our RG scheme, we now plot the inverse quasiparticle weight $Z_{\Lambda}^{-1}$ in the low-energy limit on a logarithmic scale as a function of the dimensionless ratio $U/\pi\Delta$. This is shown in Fig. 5. For comparison, we also display highly-accurate NRG data obtained from Ref. \cite{Isidori} -- which yields the correct behavior of this quantity for all coupling strengths -- and the Bethe ansatz result given by Eq. (\ref{1b}) for the case of $(U/\pi\Delta)\gtrsim 2$ (even though this latter expression is, strictly speaking, only valid in the limit $U\rightarrow \infty$). Moreover, in order to determine the boundaries as a function of the the ratio $U/\pi\Delta$ which roughly separate the weak, the moderate and the strong-coupling regimes of the model, the result predicted by conventional second-order perturbation theory (see Eq. (\ref{PT})) is also added. In this way, we observe in Fig. 5 that since perturbation theory results agree well with NRG data for $(U/\pi\Delta) <1$, the value where this dimensionless ratio is equal to unity could be viewed as marking the boundary between the weak and moderate interaction regimes. For $(U/\pi\Delta)>2.5$, the exponential (Kondo) behavior of $1/Z$ emerges, and the strong coupling regime of the model finally sets in. From all these remarks, we may conclude that our RG results for the inverse of the quasiparticle weight for the model are indeed in quantitative agreement with NRG data from weak to slightly moderate interactions (i.e., $(U/\pi\Delta)\lesssim 1.5$), which shows that the present implementation of the field-theoretical RG scheme improves somewhat the conventional perturbation theory result for this regime (see also the inset of Fig. 5). Naturally, it would be also very interesting to test our RG approach for the present model in the situation away from particle-hole symmetry, where it is well-known that the conventional perturbation theory works less well than at the particle-hole symmetric point. Indeed, in the particle-hole asymmetric case our RG approach could potentially improve even further the conventional perturbation theory result for intermediate interactions. We therefore plan to perform this analysis in a subsequent work.

On the other hand, for $(U/\pi\Delta)\gtrsim 1.5$ in the present case our RG results start to deviate quantitatively from the NRG data and the agreement with the exact results available for this model becomes only qualitative. This implies that the exponential suppression of $Z$ contained in the Bethe ansatz result is still not reproduced by our RG approach. As can be seen from Fig. 5, our results underestimate the strength of fluctuations of the SIAM for such large interactions. Therefore, this establishes the limit of validity of our two-loop truncation of the RG flow equations for describing this model at stronger couplings. Although this approximation proved to be good for not too strong interactions, it becomes evidently more and more uncontrolled for higher values of $U$. In order to obtain better results for stronger couplings in this model, it would be important to further include higher-order contributions in the self-energy (e.g., three-loop diagrams or beyond) and also in the renormalized coupling functions within the RG truncation scheme. This improvement of the results for such large interactions is indeed possible because of an important property of the SIAM, i.e. that the system always finds itself in a Fermi-liquid state for any coupling strength since it never undergoes a phase transition. In view of this fact, a perturbative truncation of the functional field-theoretical RG approach about the renormalized impurity interaction is expected to converge in this case.

We can also put our findings into context with some other works available in the literature. Karrasch \emph{et al.} recently implemented a Wilsonian functional RG study \cite{Meden} of some quantities of the SIAM such as the effective mass, the static spin susceptibility and the spectral function. They concluded that while their RG method also reproduced precise NRG results for these physical quantities at weak couplings, they failed to capture the correct Kondo scale generation for the model within their truncation. In another recent work \cite{Schoeller}, a functional RG scheme within the Keldysh formalism was implemented, and the authors obtained quantitative agreement with accurate NRG data for the SIAM in the equilibrium situation only for weak interactions. All these findings are in good agreement with our results for this regime. In addition to this, since the functional field-theoretical RG method described here turns out to be easier to implement at higher loops \cite{Shankar} than the Wilsonian approach, higher-order calculations within the present scheme are indeed feasible and could improve the results for this model at stronger couplings. In view of this, we believe the present RG scheme can be viewed as a possible alternative to other functional RG methods to describe equilibrium properties of the SIAM.

\section{Conclusion}

We have implemented a functional field-theoretical RG method up to two-loop order to the symmetric SIAM at zero temperature within the wide-band limit. One of the objectives of this study was to extend the conventional field-theoretical RG scheme to include the dynamics of all frequency-dependent renormalized parameters of the model. We have shown that, in order to accomplish this, it is convenient to define a soft ultraviolet regulator in the space of Matsubara frequencies for the renormalized Green's function of the model. In this way, we have derived analytically and then solved numerically the corresponding RG equations both for the coupling functions of the model and the quasiparticle weight in the low-energy limit. These equations fully included the interplay of particle-particle and particle-hole parquet diagrams and also the effect of two-loop self-energy feedback into them. Naturally, the RG framework was ideally suited for achieving this goal in an unbiased manner.

One important point we want to stress here is that, because of the two-loop self-energy contribution into the RG flow equations, our approach did not suffer from any instability (or artificial spin symmetry breaking) for any choice of interaction strength $U$. This is consistent with the fact that the SIAM conforms to a Fermi liquid description for all couplings at low-energies. Our RG approach therefore captured successfully this aspect of the problem. In order to assess the quality of our two-loop truncation of the RG flow equations, we have benchmarked our results against highly accurate NRG data obtained from Ref. \cite{Isidori}. As a result of this comparison, we have confirmed that our results indeed reproduce quantitatively the NRG data for the model from weak to slightly moderate interactions. Moreover, since the functional generalization of the field-theoretical RG method described here turns out to be technically easier to calculate at higher loops as compared with the more conventional Wilsonian RG approach, higher-order computations within the present scheme are relatively straightforward to implement. This latter calculation could potentially improve our results for this model at stronger couplings. Therefore, we believe the development of the present functional field-theoretical RG approach could offer a promising alternative to other functional RG methods which aim to describe electronic correlations within the Anderson impurity model.

\begin{acknowledgements}
One of us (HF) would like to thank the Brazilian funding agency CNPq through grant No. 474109/2010-0 for financially supporting this project.
\end{acknowledgements}

\appendix

\section{Appendix}

In this appendix, we write down the explicit form of all the functions defined in the text. They are given by

\begin{eqnarray}
A^{\Lambda}_{\uparrow}(\omega_{1},\omega_{2},\omega_{3})=\int_{-\infty}^{\infty}d\omega\frac{U^{\Lambda}_{\uparrow}(\omega_{1},\omega_{2},\omega)U^{\Lambda}_{\uparrow}(\omega_{1}+\omega_{2}-\omega,\omega,\omega_{3})}{(\omega+Z_{\Lambda}\Delta\,\si
(\omega))(\omega_{1}+\omega_{2}-\omega+Z_{\Lambda}\Delta\,\si
(\omega_{1}+\omega_{2}-\omega))}\mathcal{F}^{\Lambda}_{1}(\omega_{1},\omega_{2},\omega),\nonumber\\
\end{eqnarray}

\begin{eqnarray}
B^{\Lambda}_{\uparrow}(\omega_{1},\omega_{2},\omega_{3})=&&\int_{-\infty}^{\infty}d\omega\frac{1}{(\omega+Z_{\Lambda}\Delta\,\si
(\omega))(\omega_{1}-\omega_{3}+\omega+Z_{\Lambda}\Delta\,\si
(\omega_{1}-\omega_{3}+\omega))}\mathcal{F}^{\Lambda}_{2}(\omega_{1},\omega_{3},\omega)\nonumber
\\&&\Big[U^{\Lambda}_{\uparrow}(\omega_{1},\omega,\omega_{3})U^{\Lambda}_{\uparrow}(\omega_{1}-\omega_{3}+\omega,\omega_{2},\omega)+U^{\Lambda}_{\uparrow\downarrow}(\omega_{1},\omega,\omega_{3})U^{\Lambda}_{\uparrow\downarrow}(\omega_{1}-\omega_{3}+\omega,\omega_{2},\omega)\Big],
\nonumber\\
\end{eqnarray}

\begin{eqnarray}
C^{\Lambda}_{\uparrow}(\omega_{1},\omega_{2},\omega_{3})=&-&\int_{-\infty}^{\infty}d\omega\frac{1}{(\omega+Z_{\Lambda}\Delta\,\si
(\omega))(\omega+\omega_{2}-\omega_{3}+Z_{\Lambda}\Delta\,\si
(\omega+\omega_{2}-\omega_{3}))}\mathcal{F}^{\Lambda}_{3}(\omega_{2},\omega_{3},\omega)\nonumber
\\&&\Big[U^{\Lambda}_{\uparrow\downarrow}(\omega_{1},\omega+\omega_{2}-\omega_{3},\omega)U^{\Lambda}_{\uparrow\downarrow}(\omega,\omega_{2},\omega_{3})+U^{\Lambda}_{\uparrow}(\omega_{1},\omega+\omega_{2}-\omega_{3},\omega)U^{\Lambda}_{\uparrow}(\omega,\omega_{2},\omega_{3})\nonumber
\\&&-U^{\Lambda}_{\uparrow}(\omega_{1},\omega+\omega_{2}-\omega_{3},\omega)U^{\Lambda}_{\uparrow}(\omega,\omega_{3},\omega+\omega_{2}-\omega_{3})-U^{\Lambda}_{\uparrow}(\omega_{1},\omega+\omega_{2}-\omega_{3},\omega_{4})U^{\Lambda}_{\uparrow}(\omega,\omega_{2},\omega_{3})\nonumber
\\&&+U^{\Lambda}_{\uparrow\downarrow}(\omega_{1},\omega+\omega_{2}-\omega_{3},\omega)U^{\Lambda}_{\uparrow\downarrow}(\omega_{2},\omega,\omega_{3})+U^{\Lambda}_{\uparrow\downarrow}(\omega_{1},\omega+\omega_{2}-\omega_{3},\omega_{4})U^{\Lambda}_{\uparrow\downarrow}(\omega,\omega_{2},\omega_{3})\Big],
\end{eqnarray}

\begin{eqnarray}
A^{\Lambda}_{\uparrow\downarrow}(\omega_{1},\omega_{2},\omega_{3})=&-&\int_{-\infty}^{\infty}d\omega\frac{1}{(\omega+Z_{\Lambda}\Delta\,\si
(\omega))(\omega_{1}+\omega_{2}-\omega+Z_{\Lambda}\Delta\,\si
(\omega_{1}+\omega_{2}-\omega))}\mathcal{F}^{\Lambda}_{1}(\omega_{1},\omega_{2},\omega)\nonumber
\\&&\Big[U^{\Lambda}_{\uparrow\downarrow}(\omega_{1},\omega_{2},\omega)U^{\Lambda}_{\uparrow\downarrow}(\omega_{1}+\omega_{2}-\omega,\omega,\omega_{3})+U^{\Lambda}_{\uparrow\downarrow}(\omega_{1},\omega_{2},\omega)U^{\Lambda}_{\uparrow\downarrow}(\omega,\omega_{1}+\omega_{2}-\omega,\omega_{4})\Big],
\nonumber\\
\end{eqnarray}

\begin{eqnarray}
B^{\Lambda}_{\uparrow\downarrow}(\omega_{1},\omega_{2},\omega_{3})=&&\int_{-\infty}^{\infty}d\omega\frac{1}{(\omega+Z_{\Lambda}\Delta\,\si
(\omega))(\omega_{1}-\omega_{3}+\omega+Z_{\Lambda}\Delta\,\si
(\omega_{1}-\omega_{3}+\omega))}\mathcal{F}^{\Lambda}_{2}(\omega_{1},\omega_{3},\omega)\nonumber
\\&&\Big[U^{\Lambda}_{\uparrow\downarrow}(\omega_{1},\omega,\omega_{3})U^{\Lambda}_{\uparrow}(\omega_{1}-\omega_{3}+\omega,\omega_{2},\omega)+U^{\Lambda}_{\uparrow}(\omega_{1},\omega,\omega_{3})U^{\Lambda}_{\uparrow\downarrow}(\omega_{1}-\omega_{3}+\omega,\omega_{2},\omega)\Big],
\nonumber\\
\end{eqnarray}

\begin{eqnarray}
C^{\Lambda}_{\uparrow\downarrow}(\omega_{1},\omega_{2},\omega_{3})=&-&\int_{-\infty}^{\infty}d\omega\frac{1}{(\omega+Z_{\Lambda}\Delta\,\si
(\omega))(\omega+\omega_{2}-\omega_{3}+Z_{\Lambda}\Delta\,\si
(\omega+\omega_{2}-\omega_{3}))}\mathcal{F}^{\Lambda}_{3}(\omega_{2},\omega_{3},\omega)\nonumber
\\&&\Big[U^{\Lambda}_{\uparrow\downarrow}(\omega_{1},\omega+\omega_{2}-\omega_{3},\omega_{4})U^{\Lambda}_{\uparrow\downarrow}(\omega,\omega_{2},\omega_{3})+U^{\Lambda}_{\uparrow\downarrow}(\omega_{1},\omega+\omega_{2}-\omega_{3},\omega)U^{\Lambda}_{\uparrow\downarrow}(\omega,\omega_{2},\omega+\omega_{2}-\omega_{3})\nonumber
\\&&-U^{\Lambda}_{\uparrow\downarrow}(\omega_{1},\omega+\omega_{2}-\omega_{3},\omega)U^{\Lambda}_{\uparrow\downarrow}(\omega,\omega_{2},\omega_{3})\Big],
\end{eqnarray}

\noindent where

\begin{eqnarray}
&&\mathcal{F}^{\Lambda}_{1}(\omega_{1},\omega_{2},\omega)=\left(\frac{Z_{\Lambda}^{2}}{2\pi\Lambda}\right)\left[\,|\omega|+|\omega_{1}+\omega_{2}-\omega|\,\right]K(|\omega|/\Lambda)K(|\omega_{1}+\omega_{2}-\omega|/\Lambda),\nonumber
\\&&\mathcal{F}^{\Lambda}_{2}(\omega_{1},\omega_{3},\omega)=\left(\frac{Z_{\Lambda}^{2}}{2\pi\Lambda}\right)\left[\,|\omega|+|\omega_{1}-\omega_{3}+\omega|\,\right]K(|\omega|/\Lambda)K(|\omega_{1}-\omega_{3}+\omega|/\Lambda),\nonumber
\\&&\mathcal{F}^{\Lambda}_{3}(\omega_{2},\omega_{3},\omega)=\left(\frac{Z_{\Lambda}^{2}}{2\pi\Lambda}\right)\left[\,|\omega|+|\omega+\omega_{2}-\omega_{3}|\,\right]K(|\omega|/\Lambda)K(|\omega+\omega_{2}-\omega_{3}|/\Lambda).
\nonumber\\
\end{eqnarray}



\end{document}